\documentclass{WileyMSP-template}

\usepackage{xcolor}
\usepackage{graphicx, wrapfig}
\usepackage{lineno}
\usepackage{amsmath, amsfonts, mathtools}
\usepackage[font={sf}]{caption}

\usepackage{hyperref} 
\hypersetup{colorlinks=true, linkcolor=blue, filecolor=magenta, urlcolor=cyan}

\usepackage[style=numeric-comp, sorting=none, url=false]{biblatex}
\begin{document}

\pagestyle{fancy}

\title{Phononic Bandgap Programming in Kirigami}

\maketitle

\author{Hesameddin Khosravi $^*$,}
\author{Suyi Li $^+$}

\begin{affiliations}

+ Department of Mechanical Engineering, Virginia Tech\\
153 Durham Hall, 1145 Perry Street, Blacksburg, VA USA, 24060\\
\medskip
* Department of Mechanical Engineering, Clemson University\\
224 Fluor Daniel Building, 216 South Palmetto Blvd., Clemson, SC USA, 29631\\

\medskip
Correspondence: Prof. Suyi Li, \texttt{suyili@vt.edu}

\end{affiliations}


\begin{abstract}

This study investigates the programming of elastic wave propagation bandgaps in a kirigami material by intentionally sequencing its constitutive mechanical bits. Such sequencing exploits the multi-stable nature of the stretched kirigami, allowing each mechanical bit to switch between two stable equilibria with different external shapes (aka. ``(0)'' and ``(1)'' states). Therefore, by designing the sequence of (0) and (1) bits, one can fundamentally change the underlying periodicity and thus program the phononic bandgap frequencies. To this end, this study develops an algorithm to identify the unique periodicities generated by assembling ``$n$-bit strings'' consisting of $n$ mechanical bits. Based on a simplified geometry of these $n$-bit strings, this study also formulates a theory to uncover the rich mapping between input sequencing and output bandgaps. The theoretical prediction and experiment results confirm that the (0) and (1) bit sequencing is a versatile approach to programming the phonic bandgap frequencies. Moreover, one can additionally fine-tune the bandgaps by adjusting the global stretch. Overall, the results of this study elucidate new strategies for programming the dynamic responses of architected material systems.

\end{abstract}

\section{Introduction}
Advanced materials systems with on-demand property programming have enabled new capabilities for many applications. Like in computer programming, where one can design an input instruction to execute computation and obtain results, a programmable material can provide a unique output response according to an input set \cite{cui2014, liu2017}. It is worth emphasizing that these inputs should involve multiple and typically local components. So programmable materials differ fundamentally from the more traditional responsive materials that only react to a global and ambient stimulus (such as heat-responsive shape memory alloys \cite{chen2022, candido2018} or light-responsive liquid crystal polymers \cite{koccer2017, korpas2021, wu2021}). Moreover, users should be able to reset and re-design the input set for different output properties.

\medskip
Because of the local and complex nature of the input set, almost all property programming has been achieved in mechanical metamaterials consisting of an assembly of unit cells with carefully designed geometries. In particular, if each unit cell possesses two stable equilibria (aka. bi-stability), one can discretize the input space for versatile programming. Intuitively, we can refer to these bistable unit cells as ``mechanical bits'' and designate one stable configuration as the state (0) and the other as the state (1) -- similar to the binary number system in digital computation. This way, we can program the material property by inputting different combinations of (0) and (1) bits. For example, in origami-based cellular structures with $n$ bi-stable mechanical bits \cite{sengupta2018, novelino2020}, one can obtain the lowest elastic modulus if all bits are in the state (0), the highest modulus when all bits are in the state (1), and other $n-1$ levels of intermediate moduli if only portions of the mechanical bits are in the state (1).

\medskip
However, almost all material programming studies so far are limited to static responses, so the permutation (or sequence) of inputs does not play a role. In the examples above, the origami cellular material shows the same elastic modulus as long as the total number of (1) bits does not change, and the positions of these (1) bits have no influence. Therefore, we aim to explore dynamic property programming by controlling the input sets' \emph{sequences}. Specifically, we show how to program the phononic wave propagation bandgaps by sequencing the mechanical bits in a kirigami-based metamaterial.  

\medskip
Kirigami material exhibiting phononic wave bandgap is an ideal platform to investigate dynamic material property programming via input sequencing. The phononic bandgaps -- a unique phenomenon by which mechanical waves cannot propagate through periodic media due to Braggs scattering \cite{bergamini2014} or local resonance \cite{wang2004} -- is one of the most fundamental dynamic properties in a metamaterial system. The ability to manipulate wave propagation via bandgap has enabled countless applications in noise mitigation \cite{olsson2008}, wave transmission control \cite{liao2021}, acoustic cloaking \cite{chen2010}, non-reciprocity \cite{nassar2020}, and even wave-based mechanical computation \cite{zangeneh2021}. On the other hand, kirigami -- the ancient art of paper cutting -- has become a versatile engineering framework for designing and constructing multi-functional material systems \cite{tao2022}. Careful cutting can impart stretchability to thin sheet materials like metal alloys, graphene \cite{blees2015}, and electronics \cite{li2019}, thus fostering new designs of flexible sensors \cite{brooks2022}, robots \cite{rafsanjani2018, khosravi2021}, and metamaterials \cite{cho2014, bertoldi2017a, tang2017}. In particular, a stretched kirigami sheet can buckle out-of-plane and exhibit multi-stability \cite{rafsanjani2017}. One can also use simple stretching to create and control periodicity to generate tunable wave propagation bandgaps \cite{Khosravi2022}. Therefore, the results of this study can have broad applications to many engineering systems of vastly different sizes and functionalities.  

\begin{figure}[t]
    \centering
    \includegraphics[scale=1.0]{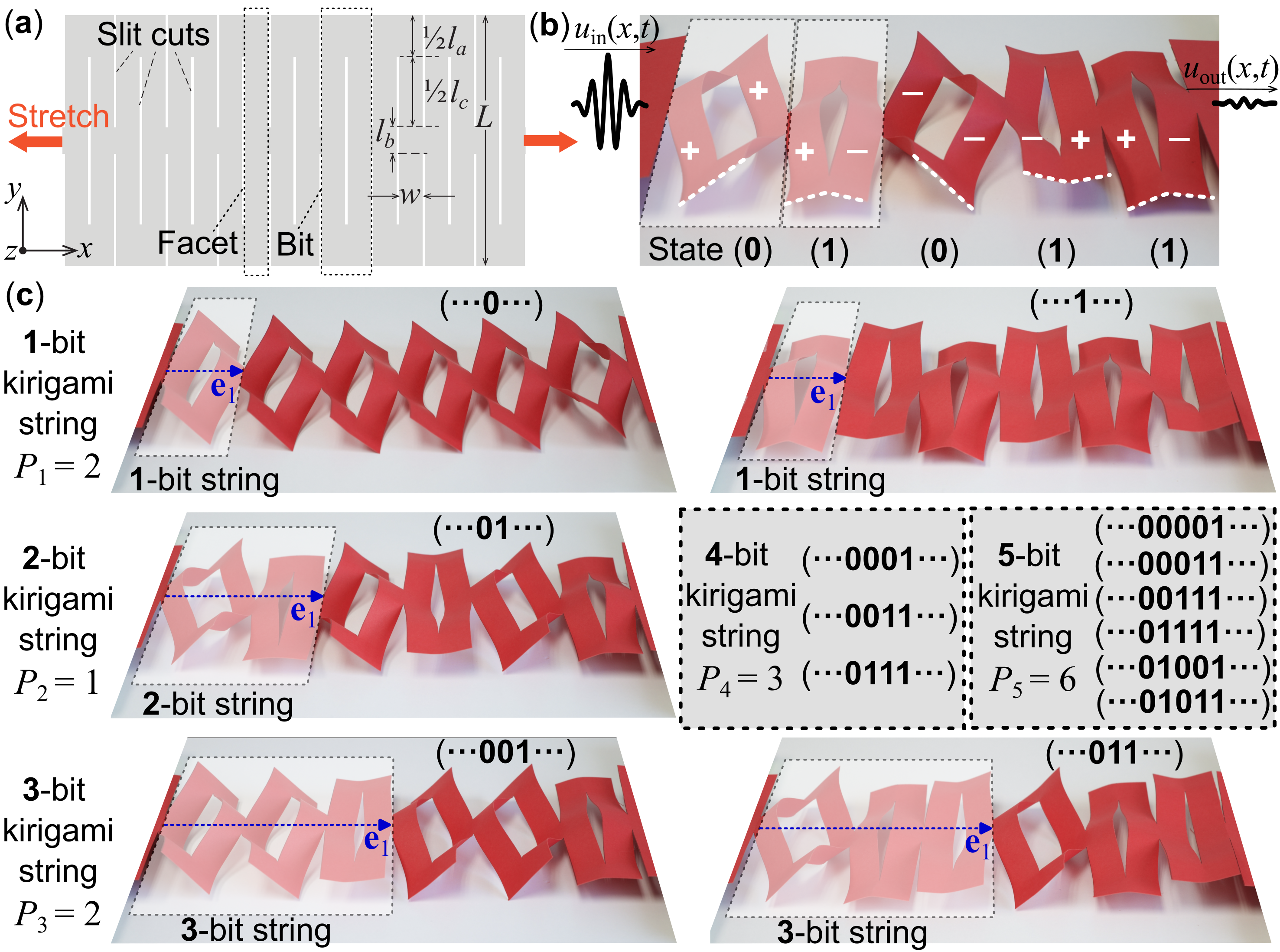}
    \caption{Design, multi-stability, and input sequencing of the stretch-buckled kirigami sheet. (a) The design parameters for a kirigami with parallel cuts. Here, the definition of facets and kirigami bits are highlighted. (b) A stretched kirigami sample containing five mechanical bits at either (0) or (1) states. The $+$ and $-$ facet rotations in each bit are highlighted for clarity. Note that the rotation direction must be the same between the two adjacent facets at the kirigami bit's boundary. Transverse elastic waves, denoted by $U(x,t)$, are sent to propagate through the kirigami sheet. (c) The unique input sequences generated by the 1, 2, and 3-bit strings, shown by a paper-based kirigami sample. The inserted gray box summarizes the unique periodicities introduced by 4 and 5-bit strings. Here, the vector $\mathbf{e}_1$ defines the fundamental periodic unit underpinning input sequence, and it is used for calculating the bandgap based on plane wave expansion method.}
    \label{fig:seq}
\end{figure}

\medskip
This study uses a simple kirigami sheet containing uniformly distributed parallel cuts (Figure \ref{fig:seq}). When this kirigami is subject to significant in-plane stretch, its slender facets could buckle and rotate out-of-plane, generating a three-dimensional periodic topology. More importantly, each elementary unit in the stretch-buckled kirigami exhibits bi-stability, so they create a linear array of mechanical bits. This setup allows us to program input -- by sequencing the (0) and (1) bits -- and fundamentally alter the underlying periodic nature. Eventually, such periodic sequencing of mechanical bits can change the elastic wave's propagation direction from longitudinal to out-of-plane, create a mechanical impedance mismatch at the junctions between adjacent bits, and generate a wave bandgap at predictable frequencies. 

\medskip
In what follows, we elucidate the working principles of (0) and (1) bit input sequencing, uncover the corresponding output bandgap programming, and validate the results via extensive experiments.

\section{Sequencing the Input}

This study focuses on the transverse elastic wave propagation in a stretched kirigami sheet with a ``zig-zag'' uniform distribution of parallel slit cuts shown in Figure \ref{fig:seq}(a).   A few simple design parameters --- including the cut size $l_a$, $l_b$, $l_c$, and the spacing between cuts $w$ --- can fully define the overall cutting pattern.  When the kirigami sheet is stretched in-plane along the direction perpendicular to the cuts (the $x-$direction), stress starts to concentrate near the cut tips, eventually forcing the facets to deform out-of-plane as the stretching reaches a critical level \cite{isobe2016}. If the kirigami sheet is stretched further after buckling, its long and slender facets will rotate accordingly. 

\medskip
Importantly, these facet rotations can occur clockwise or counter-clockwise due to the symmetric nature of kirigami (Figure \ref{fig:seq}b, for clarity, we denote the counter-clockwise facets by ``$+$,'' and clockwise ones by ``$-$'' ). In practical experimentations, the direction of facet rotation occurs randomly, dictated by fabrication imperfections and loading conditions. However, if the global stretch is constant, one could locally switch some facets' rotation directions, and the kirigami sheet can settle into the new configuration \cite{yang2018}. Therefore, the stretched kirigami sheet is multi-stable, and each stable configuration corresponds to a unique sequence of $+$ and $-$ rotated facets. 

\medskip
To better describe the programmable periodicity created by such multi-stability, we divide the stretch kirigami sheet into a linear array of bistable mechanical bits (or ``kirigami bit'' hereafter), each consisting of two slender facets (Figure \ref{fig:seq}a). A kirigami bit is at state (0) if its two facets have the same rotation direction (aka. $++$ or $--$), or state (1) if the rotation directions are opposite ($+-$ or $-+$). Although $++$ and $--$ facet pairs are mirror-symmetric to each other with respect to the $x-y$ reference plane, they have the same geometry for an incoming elastic wave. The same principle applies to the $+-$ and $-+$ pairs. 

\medskip
Our experiment efforts also reveal a critical constraint regarding the admissable facet rotation distributions: The adjacent facets' rotation direction must be the same at the boundary between two kirigami bits. For example, to construct a pair of two (0) bits (labeled as ``(00)''), one can set the underpinning facet rotations to $++ \; ++$ (or $-- \; --$). But the $++ \; --$ (or $-- \; ++$) combination turn out to be unstable in our experiment.  Likewise, for an (11) pair, its underlying facet rotation must be $-+ \; +-$ (or $+- \; -+$). The same principle applies to (01) and (10) pairs (Figure \ref{fig:seq}b). If we violate this constraint and force the two adjacent facets at the kirigami bit's boundary to rotate oppositely, the stretched kirigami sheet's neutral axis would shift locally away from the $x-y$ reference plane where the external loads are located, creating an imbalance in the resultant internal force and bending moment. 

\medskip
Then, we combine $n$ kirigami-bits into a string -- which we refer to as ``$n$-bit string'' hereafter -- and infinitely repeat such a string into a periodic topology to create the ``input sequence.'' This way, the arrangement of (0) and (1) bits in the $n$-bit string offers a mechanism for bandgap programming. The complexity of the periodicity in the input sequence is directly related to the $n$-bit string's length. For example, if the string contains only one kirigami bit ($n=1$), there are only two possible configurations: (0) or (1). So repeating a 1-bit string provides two periodic input sequences $(\cdots 0 \cdots)$ or $(\cdots 1 \cdots)$ (Figure \ref{fig:seq}c). A 2-bit string can show four possible configurations: (00), (01), (10), and (11). However, repeating (00) or (11) gives the same periodicity as those from the 1-bit string. Moreover, infinitely repeating (01) and (10) creates the same periodicity. Therefore, the 2-bit string offers one additional periodicity than the 1-bit string, so there are three unique periodic input sequences in total: $(\cdots 0 \cdots)$, $(\cdots 1 \cdots)$, and $(\cdots 01 \cdots)$. Similarly, a 3-bit string offers two new unique periodicities compared to the 2-bit string: $(\cdots 001 \cdots)$ and $( \cdots 011 \cdots)$, bringing the total number of unique input sequences to 5. 

\medskip
As $n$ increases, the number of unique periodicities increases rapidly. A generic $n$-bit string in the stretched kirigami contains a series of $n$ mechanical bits, and each can settle either in the stable state (0) or (1). Therefore, the $n$-bit string is similar to an $n$-bit binary number and has $2^n$ unique configurations (just like the $n$-bit binary number has $2^n$ unique values). However, as we just see in the 2 and 3-bit string examples, when we repeat and assemble $n$-bit strings to create an infinite periodic array as the input sequence, many show redundant periodicity. For example, a 4-bit string has 16($=2^4$) unique configurations, but it does not create 16 unique periodicities. There are two different reasons for the redundancy. 

\medskip
The first reason is that some configurations in the $n$-bit kirigami string have the same periodicity as those generated by the smaller bit strings. For example, one can configure the 4-bit string into $(0000)$. If we repeat and assemble this configuration, we will end up with an infinite array of 0 bits: $(\cdots 000000000000 \cdots)$. However, its periodicity is already considered by the 1-bit kirigami string, as shown in Figure \ref{fig:seq}(c). Similarly, $(0101)$ and $(1010)$ configurations generate the same periodicity from the 2-bit string. 

\begin{figure}[t]
    \centering
    \includegraphics[scale=1.0]{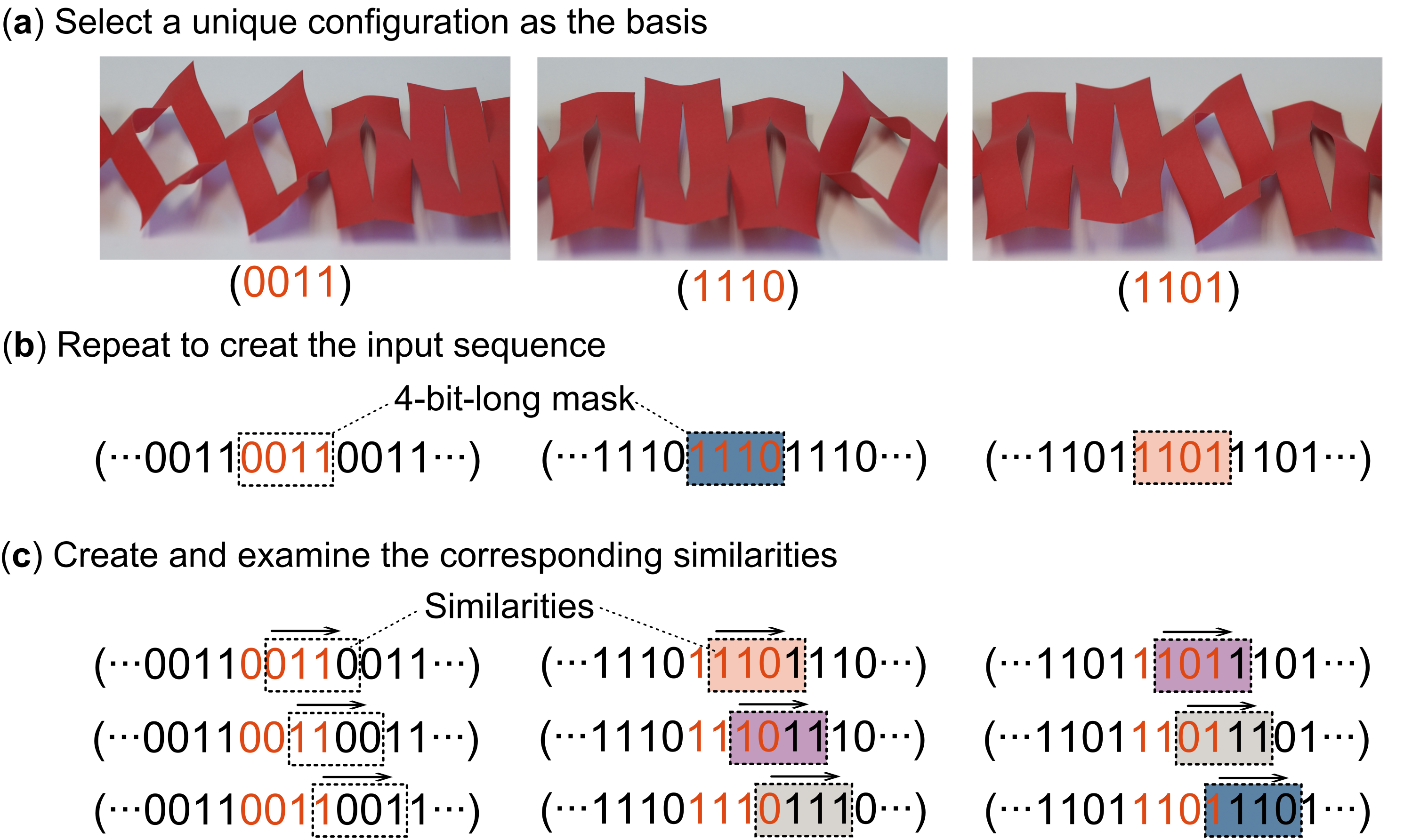}
    \caption{Identifying the redundant periodicity in the $n$-bit kirigami string, using $n=4$ as an example. (a, b) Here, we choose three unique configurations in the 4-bit string as the bases and repeat these bases to create the periodic input sequences. (c) Using a 4-bit-long mask, we make 3(=4-1) similarities for each basis by shifting the mask. For example, the similarities of the (0011) basis are (0110), (1100), and (1001). Then, we can identify redundant periodicities by searching for identical items among the similarities. In this example, the two bases (1110) and (1101) generate the same periodicity due to the many identical similarities, highlighted by the color-coded boxes.}
    \label{fig:mask}
\end{figure}

\medskip
The second reason for the redundant periodicity is that some configurations in the $n$-bit string create identical structures once they are repeated and assembled into an infinite input sequence. Again in the 4-bit string example, $(1110)$ and $(1011)$ are two different configurations. Repeating these two configurations creates $(\cdots 111011101110 \cdots)$ and $(\cdots 101110111011 \cdots)$, respectively.  However, by careful inspection, one can easily see that these two infinite arrays have the same underlying periodicity (Figure \ref{fig:mask}a-b). 

\medskip
Here, we borrow the masking and convolution concepts from image processing to search for this second type of redundant periodicity. Suppose we select a unique configuration in the $n$-bit kirigami string as the ``basis,'' we can generate $n-1$ ``similarities'' by shifting an $n$-bit-long mask successively along the corresponding infinite sequence (Figure \ref{fig:mask}c). Then, we can select another configuration as the new basis and generate its $n-1$ similarities. If identical items exist among these similarities, these two configurations are considered to create the same periodicity (Figure \ref{fig:mask}c). Moreover, since each similarity itself is also a unique configuration of an $n$-bit string, we can deduce that for each configuration of the $n$-bit kirigami string, there exist $n-1$ different configurations that could generate the identical periodic sequence. 

\medskip
Therefore, denote $P_n$ as the number of \emph{new} periodicities introduced by the $n$-bit kirigami string. Based on the discussions above, $P_1=2$, $P_2=1$, and $P_3=2$. For a $n$-bit string, we have: 
\begin{equation}
    P_n=2^n-\sum_i{iP_i}-P_n(n-1),
\end{equation} \label{eq:count}
where $2^n$ is the number of all possible configurations in the $n$-bit string. $i$ are the smaller divisors of $n$ (i.e., $i\leq n/2$), so $\sum_i{iP_i}$ corresponds to the first type of periodicity redundancy from the smaller bit strings. Here, the coefficient $i$ before ${P_i}$ includes all of the similarities in the smaller bit strings. On the other hand, $P_n(n-1)$ corresponds to the second type of redundancy mentioned above. Simplifying Equation (1) yields
\begin{equation}
    P_n=\frac{2^n-\sum_{i} i P_i}{n}, \quad \quad i=\text{divisor of } n \text{, and } i\leq \frac{n}{2}.
\end{equation} \label{eq:Pn}


\section{Mapping the Input Sequence to Output Bandgap}
Once the input sequence is programmed, the stretched kirigami takes a long-slender shape with a finite thickness in the out-of-plane $z-$direction. Therefore, we assume negligible shear deformation and rotational inertia in kirigami's cross-sections so that the classical Euler-Burnulli bean equation governs the elastic wave propagation:
\begin{equation} \label{eq:eom}
    \frac{\partial^2}{\partial x^2} \left[E I(x) \frac{\partial^2 {U(x,t)}}
    {\partial x^2}\right]+ \rho A(x)\frac{\partial^2 {U(x,t)}}{\partial t^2} = 0,
\end{equation}
where $\rho$ and $E$ are the mass density and Young's modulus of the constitutive sheet material, respectively. $U(x,t)$ is the out-of-plane transverse wave displacement field in the $z-$direction. $A(x)$ and $I(x)$ are the spatial distribution of cross-section area and area moment of inertia, respectively. Note that the in-plane stretching force does not play any significant role here because right after the stretch-induced buckling occurs, the kirigami's reaction force does not increase significantly unless the stretch becomes high enough to generate plastic deformation \cite{rafsanjani2017}. 

\begin{figure}[t]
    \centering
    \includegraphics{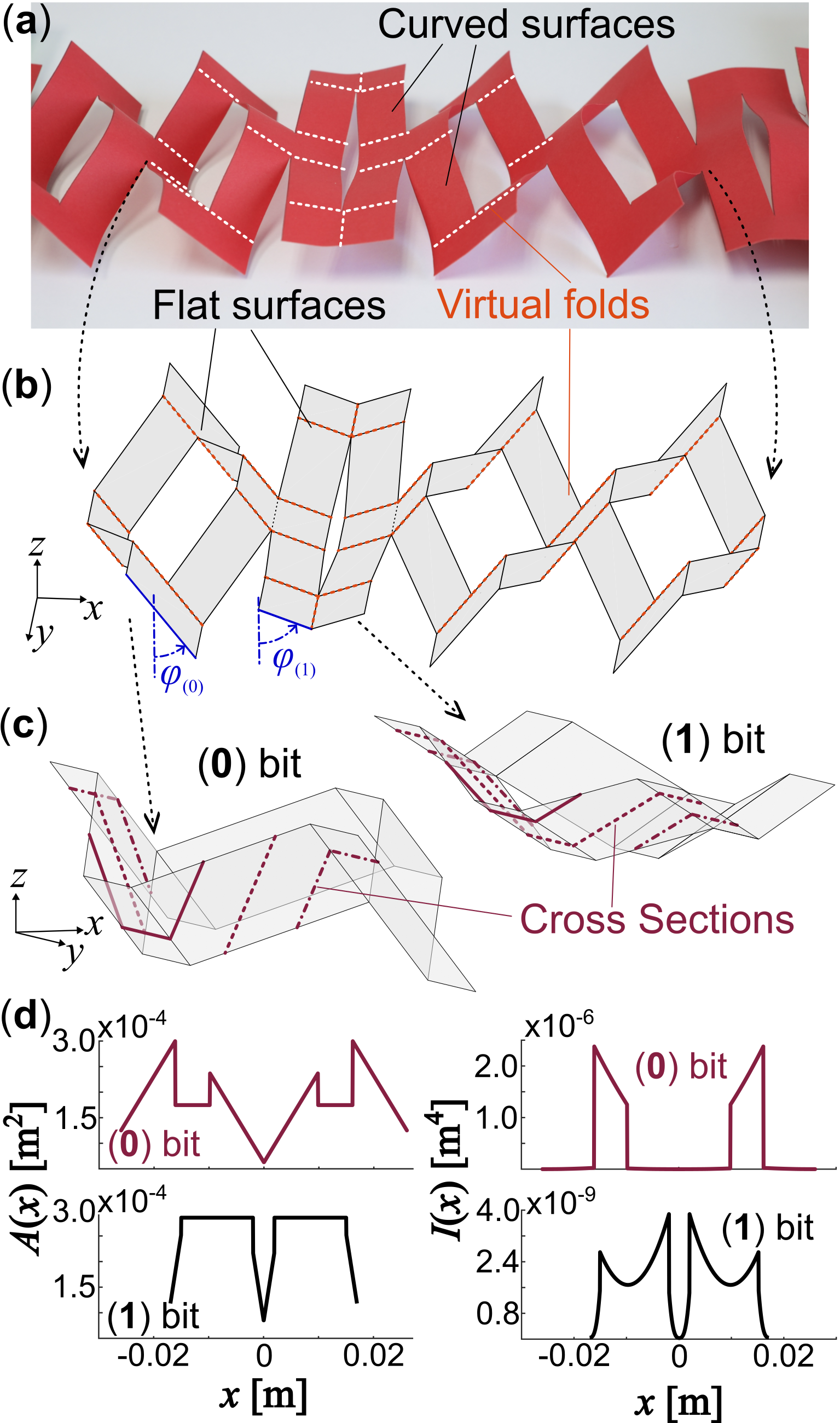}
    \caption{Geometric simplification of stretched kirigami (a) A close-up view of a stretched kirigami, showing its complex curved facets. (b) The corresponding simplified geometry using the virtual folds and flat surface assumptions. Notice the definition of $\varphi$ angle. (c) A detailed view of the simplified (0) and (1) bits. Here, the straight cross-sections at different positions in the kirigami bit are highlighted. (d) Examples of $A(x)$ and $I(x)$ distributions ($l_a=0.044$m, $l_b=0.022$m, $l_c=0.086$m, $w=0.016$m, $\varphi_\text{(0)}=0.66$, and $\varphi_\text{(1)}=1.23$).}
    \label{fig:geometry}
\end{figure}

\medskip
Calculating $A(x)$ and $I(x)$ is quite challenging because the kirigami facets have complex-curved shapes orientated at an angle with respect to the neutral $x-y$ plane. Therefore, we adopt the simplification method proposed in the previous literature that uses virtual origami folds and flat surfaces to approximate the stretch-buckled kirigami's geometry \cite{Khosravi2022} (Figure \ref{fig:geometry}). This method has been proven successful in providing reasonably accurate predictions of bandgap frequency without incurring a high computational cost. 

\medskip
According to the geometric simplification, the (0) bit consists of 8 flat surfaces and 8 fold lines, and all these fold lines are perpendicular to the slit cuts. On the other hand, a (1) bit has 2 additional fold lines aligned with the cuts (Figure \ref{fig:geometry}). In this way, the post-buckling deformation of the kirigami becomes a rigid-folding motion with one kinematic degree of freedom. To this end, we choose the dihedral angle ($\varphi$) between the rotated facets in the kirigami bit and the $y-z$ reference plane as the independent variable (Figure \ref{fig:geometry}b). When the kirigami sheet is un-deformed (or flat in the $x-y$ reference plane), $\varphi$ takes the maximum value ($\varphi_{max} = \pi/2$); when the buckled kirigami sheet is fully stretched, $\varphi$ take the minimum value:
\begin{equation}
    \varphi_{min} = \tan^{-1} \left( \frac{2w}{l_c} \right).
\end{equation}

The length of the kirigami mechanical bits is:
\begin{equation}
    a = \frac{2w}{\sin\varphi}.
\end{equation}

Using the geometric simplification, one can represent the in-plane stretching stiffness of the post-bucked kirigami bits by assuming that the virtual folds behave like hinges with embedded torsional springs and that the surface between them is rigid \cite{li2019a}. Intuitively, (1) bit is stiffer in stretching than (0) bit because it has two additional virtual folds (thus two additional torsional springs). Therefore, (1) bit is shorter than a (0) bit along the in-plane $x-$direction because they are mechanically connected in series. We stretched a kirigami sample containing four (0) bits and four (1) bits to different global lengths and measured the kirigami bits' lengths locally, and found a simple linear relationship:
\begin{equation}
    a_{(1)} =  \frac{a_{(0)}+33 \text{mm}}{2.53} \pm 0.43\text{mm}.
\end{equation}

Moreover, the cross-sections of simplified kirigami bits are straight lines (Figure \ref{fig:geometry}c), so the calculation of $A(x)$ and $I(x)$ becomes manageable. Figure \ref{fig:geometry}(d) summarizes the results corresponding to the (0) and (1) bits. Section 1 of the supplement materials details the underpinning formulations.

\medskip
To analyze the wave propagation behavior in the stretch-buckled kirigami sheet, we employ the plane wave expansion (PWE). First, we define $\mathbf{R} = a_1 \mathbf{e}_1$ as a translational vector in the \emph{direct space} to represent the periodicity of the programmed kirigami sheet. In other words, we can construct the kirigami sheet by a set of infinite translational operations on the $n$-bit string based on $\mathbf{R}$. In this case, $a_1$ is an integer number, and the lattice vector $\mathbf{e}_1$ connects the two ends of the $n$-bit string (Figure \ref{fig:seq}c). We then define the reciprocal lattice vector $\mathbf{G}$ in that $e^{-i\; \mathbf{G} \cdot \mathbf{R}}=1$. Here, $m_1$ is another integer, $\mathbf{G}= m_1 \mathbf{b}_1$, and $\mathbf{b}_1$ is the lattice vector in the \emph{reciprocal space}.  Equivalent to the lattice vector $\mathbf{e}_1$ in the direct space, infinite translation operations of the reciprocal lattice vectors $\mathbf{b}_1$ defines the \emph{reciprocal space}, which has similar symmetry as the direct space. Like in the direct lattice, an equivalent periodic unit can then be defined in the reciprocal space based on $\mathbf{b}_1$, giving us the First Irreducible Brillouin Zone (FBZ).

\medskip
Then, a separation of variables gives $U(x,t)=\tilde{U}(x)e^{-i\omega t},$ where $
\omega$ is the harmonic oscillation frequency.  Based on the Blotch theorem, we can formulate the spatial term $\tilde{U}(x)$ for the whole kirigami sheet into a product of plane-wave and periodic functions in the First Irreducible Brillouin Zone so that
\begin{equation}
    \tilde{U}(x)= e^{i (k_1 b_1 x)} \sum_{n_1}{\hat{U}({n_1b_1})} e^{i(n_1 b_1 x)},
\end{equation}
where $b_1$ is the magnitude of reciprocal lattice vector ($b_1=|\mathbf{b}_1|$), $n_1$ is an integer, and ${k}_1 $ is a wave number within the FBZ in that ${k}_1 \in [0,\frac{1}{2}]$. We further expand the distribution of cross-section area $A(x)$ and area moment of inertia $I(x)$ of an $n$-bit string into Fourier series so that
\begin{align}
    I(x) & = \sum_{m_1}\hat{I}(m_1 b_1)e^{i(2\pi m_1 x)},\\
    A(x) & = \sum_{m_1}\hat{A}(m_1 b_1)e^{i(2\pi m_1 x)}.
\end{align}

By substituting the formulations above into the governing equations of motion (\ref{eq:eom}), we obtain the eigenvalue problem for calculating the dispersion curves and bandgap frequencies. 
\begin{equation}
    \sum_{n_{1}} E(n_1+k_1)^2(k_1+m_1+n_1)^2\hat{I}(m_1 b_1) - \omega^2\rho \hat{A}(m_1b_1)\;= 0.
\end{equation}

Section 2 of the supplement material summarizes the dispersion curves from 1, 2, and 3-bit strings, and interested readers can refer to the authors' previous publication for a more detailed formulation underpinning the PWE method \cite{Khosravi2022}.

\section{Programming and Fine-Tuning the Output Bandgap}

To validate the mapping between input programming and output bandgap, we compare the theoretical prediction to experiment measurements based on a Nylon-based kirigami sample with 13 mechanical bits ($E=5.09$GPa, $\rho=1200$Kg/m$^3$, material thickness $t=0.00175$m, with cut sizes $l_a=0.044$m, $l_b=0.022$m, $l_c=0.086$m, and cut spacing $w=0.016$m). Figure \ref{fig:ExpSetup} in Section 3 of the supplement material details the experiment setup. We attach one end of the kirigami sheet to a shaker (Labworks DB-140 with Pa-141 amplifier) to provide the out-of-plane ($z-$direction) excitation. The other end is attached to a rigid end fixture via thin tape to mimic the free boundary conditions. To measure the wave transmissibility through the stretch-buckled kirigami sheet, we use a signal generator (Tektronix AFG3022c) to generate harmonic excitations with sweeping frequencies and use two laser vibrometers (Polytec OFC-5000) to measure the input and transmitted waves at the two ends of kirigami. Then, we calculate the wave transmissibility, defined as $TR= \tilde{U}_\text{out} / \tilde{U}_\text{in}$, where $\tilde{U}_\text{in}$ and $\tilde{U}_\text{out}$ are the maximum displacement of incoming and transmitted wave, respectively. This transmissibility has been used widely to evaluate bandgap frequency \cite{Khosravi2022}.  It is worth noting that as we change the sequence of input (0) and (1) bits, we also adjust the global stretch of the kirigami so that the length of (0) and (1) bits remain the same among different tests: $a_{(0)}=0.052$m, $a_{(1)}=0.034$m.

\begin{figure}[t]
    \centering
    \includegraphics[scale=1.0]{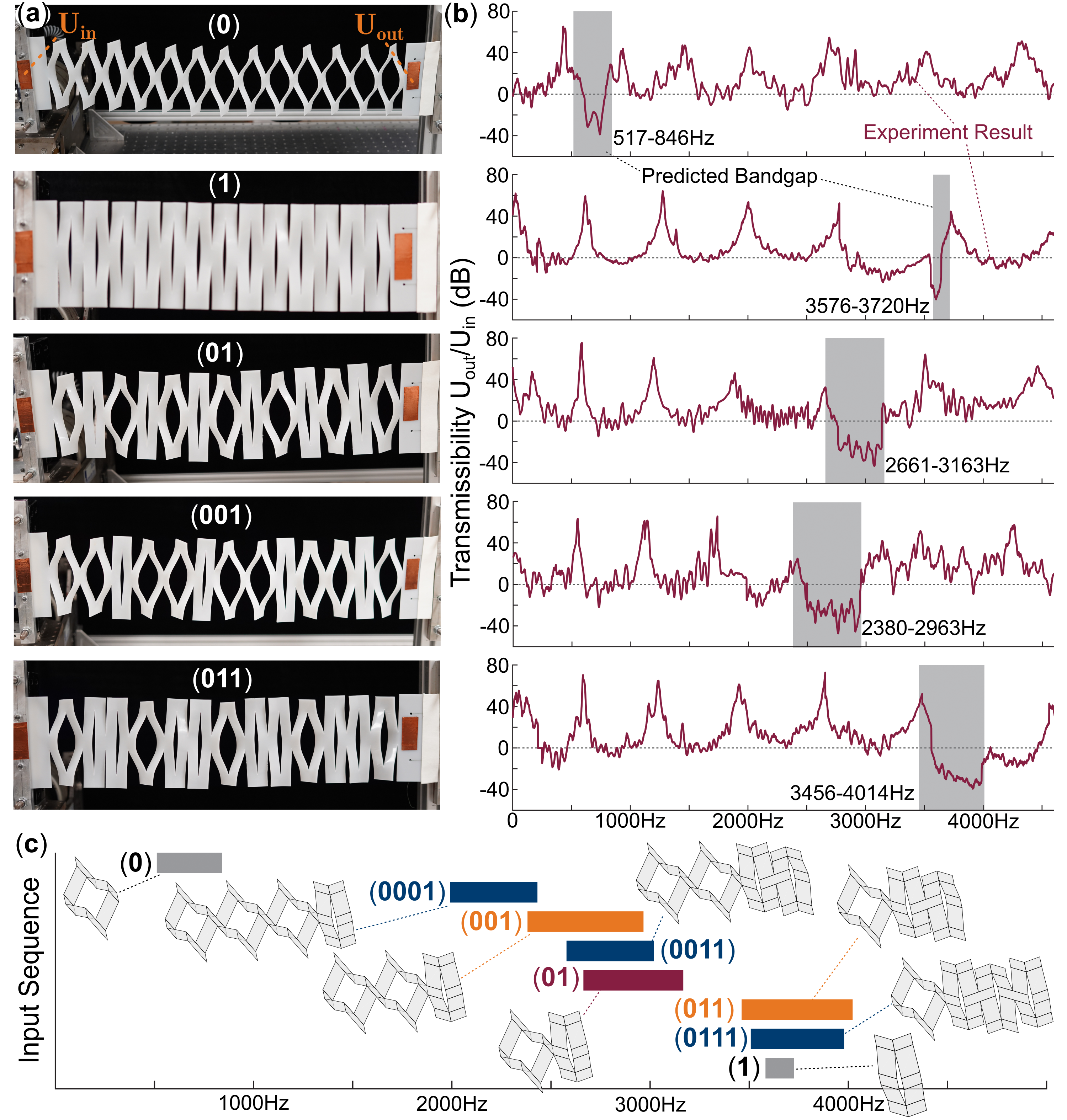}
    \caption{Phononic bandgap programming in kirigami by input sequencing. (a) Test results from 1, 2, and 3-bit strings validate the bandgap programming conectp, with good agreement between theoretically predicted bandgaps and experimentally measured wave transmissibility. (b) Summary of theoretically predicted bandgaps of 1, 2, 3, and 4-strings, arranged according to the percentage of their mechanical bits in the state (1). That is, the input sequence at the top has the most (0) bits, and the one at the bottom has the most (1) bits.}
    \label{fig:programming}
\end{figure}

\medskip
The analytical theory's predictions and experimental results all confirm the occurrence of Bragg's bandgap in the stretch buckled kirigami sheet (Figure \ref{fig:programming}). More importantly, these bandgap frequencies can be effectively programmed by simply sequencing the (0) and (1) bits. Based on the theoretical prediction, if we use the 1-bit string to sequence the input, the first bandgap occurs at 517 to 846Hz when the 1-bit string is (0) (aka. all kirigami bits are in state 0). However, the bandgap jumps to a much higher frequency at 3576 to 3720Hz when the 1-string is (1). By using the 2-bit string (01) to sequence the input, we can achieve a new bandgap at 2661 to 3163Hz. Similarly, the 3-bit string offers two additional unique periodicities compared to the 2-bit string (aka. $P_3=2$), and we obtained two new bandgaps accordingly (Figure \ref{fig:programming}a). Figure \ref{fig:programming}(b) summarizes the theoretically predicted bandgap frequencies up to the 4-bit string. The bandgap generally moves to higher frequencies and becomes narrower when more kirigami bits are in the state (1). However, if the total number of (1) bits stays the same, the bandgap would jump to different frequencies as we change the sequence of these (1) bits, elucidating the rich programmability of the bandgaps in kirigami. For example, the 2-bit (01) and 4-bit (0011) both have 50\% of their kirigami bits in the state (1), and they show different bandgap frequencies due to their unique sequencing. Our experiment results based on 1, 2, and 3-bit strings agree well with the theoretical prediction, validating the bandgap programmability.

\medskip
While the experiment results and theoretical predictions show good agreement regarding the bandgap frequencies for all input sequences, some minor differences still occur. These discrepancies come from a combination of different factors. For example, the simplified geometry used in our theoretical prediction approximates the actual shapes of the stretch-buckled kirigami bits, so it would inevitably introduce errors, especially for the (1) bits. Moreover, the analytical model based on the plane wave expansion method assumes an infinite number of mechanical bits in the kirigami sheet, while the experiment prototype only has thirteen bits. The omission of the tension force in the governing equation is another probable cause for discrepancies. Regardless, these results firmly validate our assumption that by sequencing the (0) and (1) mechanical bits, one can program the elastic wave propagation bandgaps with considerable freedom. Moreover, our analytical model elucidates the physical principles underpinning the mapping between input sequencing and output bandgap frequency.

\begin{figure}
    \centering
    \includegraphics[scale=1.0]{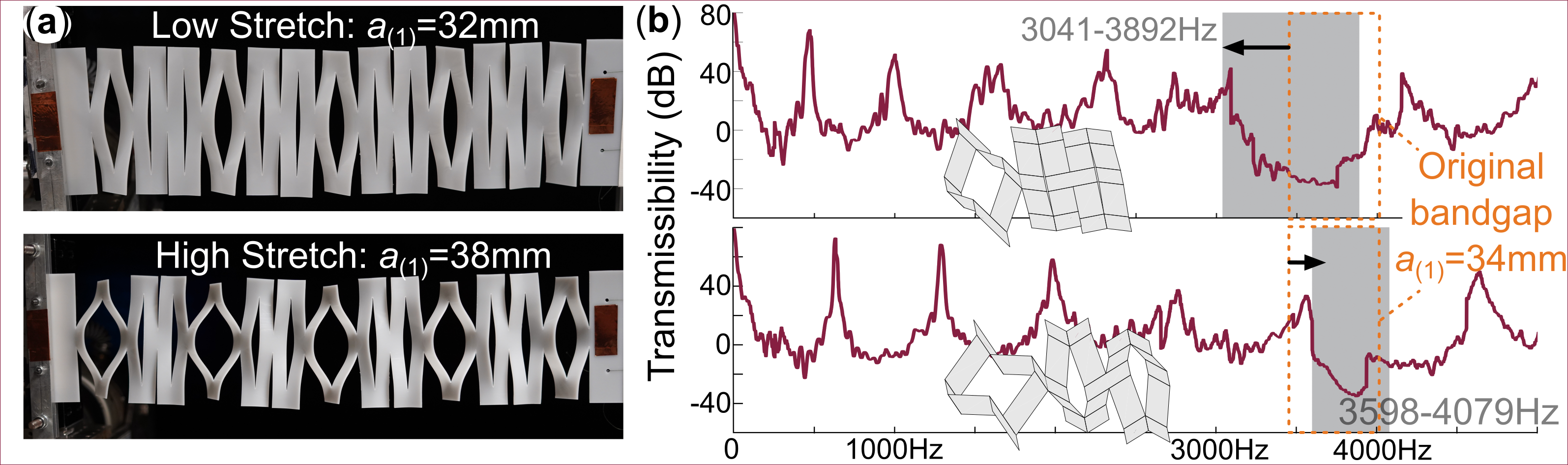}
    \caption{Fine-tuning the bandgap by adjusting the global external stretch. By lowering the global stretch from $a_{(1)} = 0.034$m to  $a_{(1)} = 0.032$m, or increasing the stretch so that $a_{(1)} = 0.038$m, one can continuously adjust the bandgap frequencies.}
    \label{fig:Tuning}
\end{figure}

\medskip
The input sequencing of (0) and (1) kirigami bits offer an avenue to program the phononic bandgaps to different frequency ranges. In addition, we can fine-tune the bandgap by slightly adjusting the global stretch. In other words, if we consider the (0) and (1) bit sequencing as a ``digital'' and discrete method to program bandgap frequency, adjusting the kirigami stretch is an ``analog'' and continuous approach to fine-tune bandgaps. In a follow-up experiment, we adjust the global stretch for the kirigami sheet -- based on a 3-string (011) input sequence -- and continuously move the bandgap frequency (Figure \ref{fig:Tuning}). Therefore, to achieve a targeted phononic bandgap frequency, one can first use the (0) ad (1) bits input sequencing to program the bandgap near the target and then adjust the global stretching to reach the target precisely. 

\section{Summary}
This study theoretically and experimentally examines the phonic bandgap programming in stretched kirigami material with intentionally sequenced mechanical inputs. The kirigami consists of a linear array of mechanical bits, each exhibiting two distant stable equilibria due to the internal stress generated by stretching. We refer to them as (0) and (1) states, inspired by the binary number systems underpinning computer programming. Therefore, by designing the sequence of these (0) and (1) bits, one can fundamentally change the underlying periodicity in kirigami and thus shift the phononic bandgap frequencies. We develop an algorithm to identify the unique periodicities generated by repeating and assembling a string of $n$ kirigami bits. An analytical theory is also formulated to uncover the rich mapping between input sequencing and output bandgaps. This theory leverages a simplified geometry of stretched kirigami using the virtual fold and flat surface assumption and incorporates the plane wave expansion method to calculate the bandgap structures. Our theoretical prediction and experiment results confirm that (0) and (1) bit sequencing is a versatile and effective approach to programming the phonic bandgap frequencies. Moreover, one can additionally fine-tune the bandgaps by adjusting the global stretch. Overall, the results of this study elucidate new approaches to programming the dynamic responses of architected material systems.

\medskip
\textbf{Acknowledgements} \par
The authors acknowledge the support from National Science Foundation (CMMI-2240211 CAREER, 2240326).

\printbibliography

\newpage
\section*{Supplement Material}

\subsection*{1. Simplified Geometry of Kirigami Bits}

To analytically investigate the band gap distribution and its correlation to input sequencing, we simplify the kirigami sheet geometry by assuming that each mechanical bit consists of flat surfaces connected by virtual fold lines (Figure \ref{fig:geometry}). In this way, the post-buckling deformation of the kirigami becomes a rigid-folding motion with one kinematic degree of freedom.  

\begin{figure}
    \centering \includegraphics[scale=0.9]{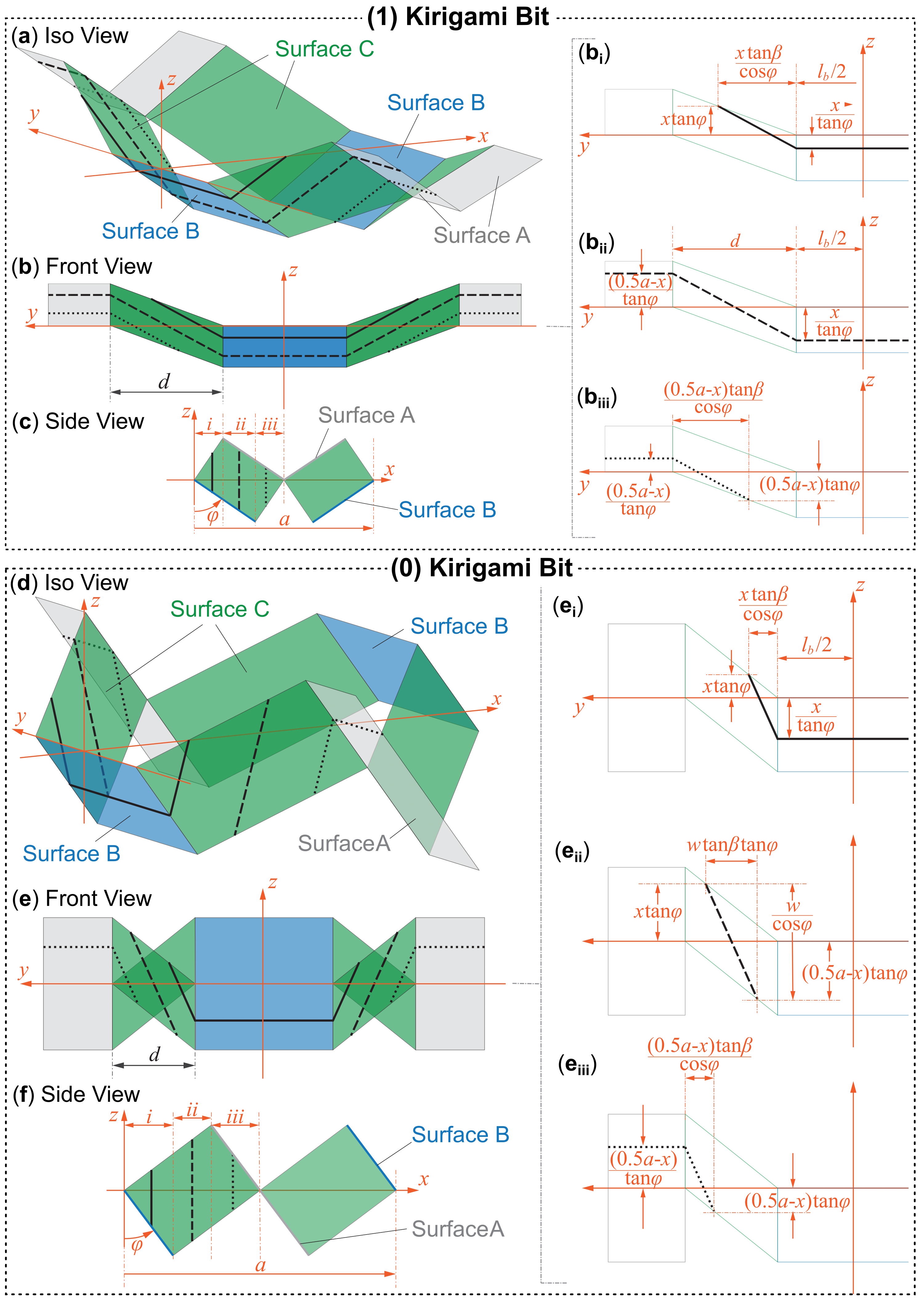}
    \caption{The simplified geometry of (1) and (0) kirigami bits. (a-c): Isometric, front, and side views of the (1) bit, respectively. (b$_\text{i}$-b$_\text{iii}$): Close-up front views of the cross-section areas corresponding to different $x$ values. Notice the solid, dashed, and dotted black lines are the cross-section area of the (1) bit in the three different sections. (d-f) The different views of a (0) bit, where the closed-up front views of its cross-section areas are detailed in (e$_\text{i}$-e$_\text{iii}$).}
    \label{fig:bit}
\end{figure}

\medskip
Figure \ref{fig:bit} details the simplified geometry of a kirigami bit. We choose the dihedral angle ($\varphi$) between surface B and the $y-z$ reference plane as the independent variable.  When the kirigami sheet is un-deformed (or flat in the $x-y$ reference plane), $\varphi$ takes the maximum value ($\varphi_{max} = \pi/2$); when the buckled kirigami sheet is fully stretched, $\varphi$ take the minimum value:
\begin{equation}
    \varphi_{min} = \tan^{-1} \left( \frac{2w}{l_c} \right).
\end{equation}

We also denote the dihedral angle between surface C defined in Figure \ref{fig:bit} and the $x-z$ reference place as $\beta$, so that

\begin{equation}
    \beta = \cos^{-1} \left(\frac{2w}{l_c \tan \varphi} \right).
\end{equation}

Clearly, $\beta = \pi/2$ when the kirigami bit is un-deformed (flat), and $\beta=0$ when the Kirigami sheet is fully stretched. The overall length of the unit cell is

\begin{equation}
    a = \frac{2w}{\sin\varphi}.
\end{equation}

Another important geometric variable is the \emph{projected} length of surface C along the $y$ axis as illustrated in Figure \ref{fig:bit}(b, e):

\begin{equation}
    d =w\frac{\tan \beta}{\tan \varphi} =\sqrt{\frac{l_c^2}{4}-\frac{w^2}{\tan^2 \varphi}}=\frac{l_c}{2}\sin \beta.
\end{equation}

Based on the geometric variables defined above, we can derive the cross-section area $A(x)$ and the corresponding area moment of inertia $I(x)$ of the kirigami bit. Here, we discuss two different cases for (1) bit and (0) bit.

\medskip
{\bf Case 1. (1) bit:} The first case corresponding to the slightly stretched Kirigami sheet after buckling in that $\pi/4<\varphi<\pi/2$ (assuming $l_c>2W$). We can divide half of the mechanical bit into three sections because the cross-section areas take distinct shapes in these three sections (Figure \ref{fig:bit}c).

\medskip
The first section corresponds to $0 < x < a/2-W \sin \varphi$, and the cross-sectional areas involves surface B and surface C (solid lines in Figure \ref{fig:bit}b$_i$).  The cross-section area is the summation of these two parts so that

\begin{equation}
    A_i(x) =\frac{t}{\sin \varphi}\left(l_{b}+2L_{i}^C\right)=\frac{t}{\sin \varphi} \left(l_b + \frac{x}{w \sin \varphi} \sqrt{4w^2+l_c^2 \tan^4 \varphi} \right),
\end{equation}
where $t$ is the thickness of the Kirigami sheet material.  The bending moment of inertia {\it with respect to the $y$-axis} also includes two parts:

\begin{equation}
    I_i(x) = I^{B}+l_b\frac{t}{\sin \varphi}\left(\frac{x}{\tan \varphi}\right)^2+2 I_{i}^C + 2\frac{x \tan \beta}{\cos \varphi}\frac{t}{\sin \varphi}\left(\frac{x\tan \varphi - x\cot \varphi}{2}\right)^2,
\end{equation}
where the first two terms come from the cross-section in surface B (using parallel axis theorem), and the second two terms come from the cross-section in surface C. Here, $I^B$ is the bending moment of inertia of the surface B cross-section with respect to its own center of mass in that

\begin{equation}
    I^{B} = \frac{l_b}{12}\left(\frac{t}{\sin \varphi} \right)^3.
\end{equation}

Similarly, $I_i^C$ is the bending moment of inertia with respect to its own mass center in that

\begin{equation}
    I_i^C(x) = \frac{t}{12 \sin \varphi} \frac{x \tan \beta}{\cos \varphi} \left[ \frac{t^2}{\sin^2 \varphi}+ x^2(\tan \varphi+\cot \varphi)^2\right].
\end{equation}

The second section corresponds to $a/2-w \sin < x < w \sin \varphi$, and the Kirigami cross-section area here involves surface A, B and C.

\begin{equation}
    A_{ii}(x) = \frac{t}{\sin \varphi} \left(l_{a}+l_{b}+2 L_{ii}^C \right) =\frac{t}{\sin \varphi}\left(l_a + l_b + \sqrt{\frac{4w^2}{\tan^4 \varphi}+l_c^2} \right).
\end{equation}

The bending moment of inertia {\it with respect to the $y$-axis} includes three parts:

\begin{equation}
    \begin{split}
    I_{ii}(x) & = I^{A}+l_a\frac{t}{\sin \varphi}\left(\frac{a/2-x}{\tan \varphi}\right)^2+ I^{B}+l_b\frac{t}{\sin \varphi}\left(\frac{x}{\tan \varphi}\right)^2 \ldots \\
    & +2 I_{ii}^C + 2 d\frac{t}{\sin \varphi}\left(\frac{a/4-x}{\tan \varphi}\right)^2.
    \end{split}
\end{equation}

Here, $I^A$  is the bending moment of inertia of the cross-section area in surface A with respect to its own mass center so that

\begin{equation}
    I^{A} = \frac{l_a}{12}\left(\frac{t}{\sin \varphi} \right)^3,
\end{equation}
and 

\begin{equation}
    I_{ii}^C=\frac{t}{12 \sin \varphi}\frac{W \tan \beta}{\tan \varphi}\left(\frac{t^2}{\sin^2 \varphi} + \frac{a^2}{4 \tan^2 \varphi}\right).
\end{equation}

Finally, the third section of the half unit cell corresponds to $\sin \varphi <x< a/2$.  In this section, the cross-section area involve surface A and C in that

\begin{equation}
    A_{iii}(x) = \frac{t}{\sin \varphi} \left(l_{a}+2 L_{iii}^C \right) =\frac{t}{\sin \varphi}\left(l_a + \frac{a/2-x}{W \sin \varphi} \sqrt{4W^2+l_c^2 \tan^4 \varphi} \right)
\end{equation}

The corresponding bending moment of inertia with respect to the $y-$axis includes two components in that,  

\begin{equation}
    \begin{split}
    I_{iii}(x) & = I^{A}+l_a\frac{t}{\sin \varphi}\left(\frac{a/2-x}{\tan \varphi}\right)^2 \ldots \\
    &+2 I_{iii}^C + 2\frac{(a/2-x) \tan \beta}{\cos \varphi}\frac{t}{\sin \varphi}\left[\frac{(a/2-x)(\tan \varphi - \cot \varphi)}{2}\right]^2.
    \end{split}
\end{equation}
where, 

\begin{equation}
    I_{iii}^C(x) = \frac{t}{12 \sin \varphi} \frac{(a/2-x) \tan \beta}{\cos \varphi} \left[ \frac{t^2}{\sin^2 \varphi}+ (a/2-x)^2(\tan \varphi+\cot \varphi)^2\right].
\end{equation}

\medskip
{\bf Case 2. (0) bit:} This case corresponds to $\varphi \in [\varphi_{min} \ldots \pi/4]$ (Figure \ref{fig:bit}d-f). Again, we can divide half of the kirigami bit into three different sections based on the shape of its cross-section. Here, the cross-sectional area $A(x)$ and bending moment of inertia $I(x)$ in the first section $(0< x <w\sin \varphi)$ and third section $(a/2-w \sin \varphi<x<a/2)$ are the same as those in the previous case. The only difference is in section ii, where the cross-section only involves surface C so that,

\begin{equation}
    A^*_{ii}(x) = 2 \frac{t}{\sin \varphi} L_{ii}^{C*}  =\frac{t}{\sin \varphi} \sqrt{4w^2+l_c^2 \tan^4 \varphi},
\end{equation}
and

\begin{equation}
    I^*_{ii}(x)  = 2 I_{ii}^{C*} + 2 \frac{t}{\sin \varphi}\left[(a/4-x)\tan \varphi \right]^2,
\end{equation}
where

\begin{equation}
    I_{ii}^{C*} =\frac{t}{12\sin \varphi} w \tan \varphi \tan \beta \left( \frac{t^2}{\sin^2 \varphi}+\frac{w^2}{\cos^2 \varphi} \right)
\end{equation}

\newpage

\subsection*{2. Dispersion Relationship Based on 1, 2, and 3-Bit Strings}

\begin{figure}[h!]
    \centering \includegraphics[scale=1.0]{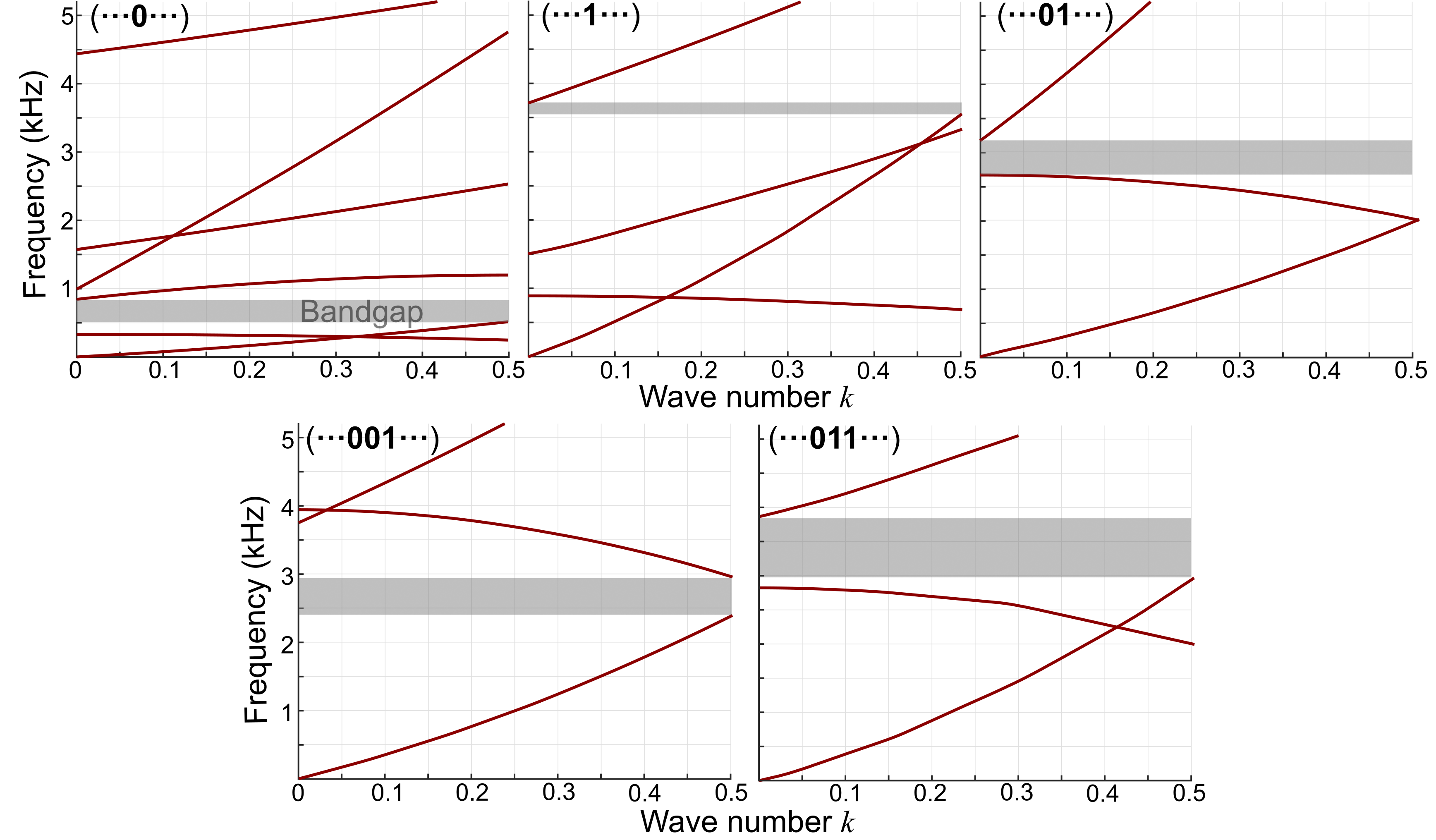}
    \caption{The dispersion relationship of kirigami sheets based on 1, 2, and 3-bit strings. The underpinning kirigami design and material properties are summarized at the beginning of Section 4.}
    \label{fig:dispersion}
\end{figure}

\newpage
\subsection*{3. Experiment Setup}
\begin{figure}[t]
   \centering
    \includegraphics[scale=0.15]{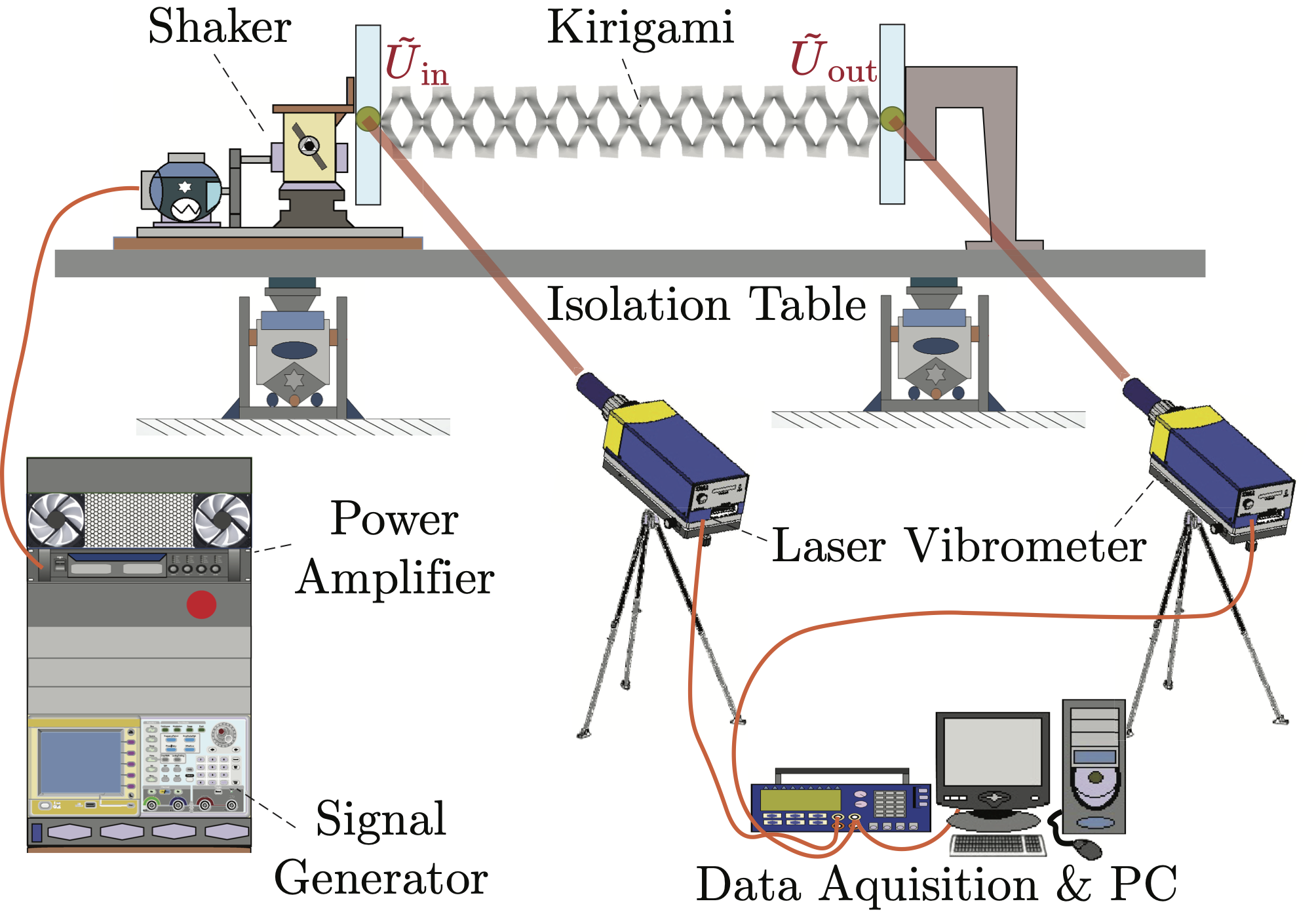}
    \caption{A schematic diagram of the test apparatus.}
    \label{fig:ExpSetup}
\end{figure}

Figure \ref{fig:ExpSetup} details the complete experiment setup for measuring the transverse elastic wave transmissibility over the stretch-buckled kirigami.  It is worth noting that the unit cells in Figure\ref{fig:seq} c  appear different from each other, but this is an optical illusion.   All unit cells in the kirigami prototype, except for the first and last one at the boundary, have the same shape and dihedral angle.  However, when the camera is placed in front of the kirigami prototype, every unit cell is at a slightly different angle with respect to the camera lens, so they appear different in the picture.  

\medskip
Moreover, we did not observe significant global plastic deformation in our stretched kirigami sample because the zig-zag distributed parallel cuts are intended to offer high stretchability and elastic response.  Due to stress concentration, there are small and local plastic deformations at the tip of the slit cuts.  These plastic deformations can be avoided by optimizing the cut tip shape, and they should not significantly affect the elastic wave bandgaps at low frequencies.


\end{document}